\documentstyle[twocolumn,prb,aps]{revtex}
\begin{document}
\draft

\twocolumn[
\hsize\textwidth\columnwidth\hsize\csname@twocolumnfalse\endcsname

\preprint{To Appear in \prb (Rap.\ Comm.)}
\title{Antiferromagnetic Excitations
and Van Hove Singularities in YBa$_2$Cu$_3$O$_{6+x}$}
\author{
G.~Blumberg,$^{1,2}$ Branko\ P.\ Stojkovi\'c$^{1,3}$, and
M.\ V.\ Klein$^{1}$
}
\address{NSF Science and Technology Center for Superconductivity
and\\
$^{1}$Department of Physics,
University of Illinois at Urbana-Champaign, Urbana, IL
61801-3080\\
$^{2}$Institute of Chemical Physics and Biophysics, R\"avala 10,
Tallinn EE0001, Estonia\\
$^{3}$Minnesota Supercomputer Institute, Minneapolis, MN 55455}
\date{September 20, 1995; to appear in \prb (Rap.\ Comm.)}
\maketitle

\begin{abstract}
We show that in quasi-two-dimensional $d$-wave superconductors
Van Hove singularities close to the Fermi surface lead to novel
magnetic
quasi-particle excitations.
We calculate the temperature and
doping dependence of dynamical magnetic susceptibility for YBCO
and
show that the proposed excitations are in
agreement with inelastic neutron scattering experiments.
In addition, the values of the gap parameter and in-plane
antiferromagnetic coupling are much smaller than usually believed.
\end{abstract}

]

The spin dynamics of
the cuprate superconductors is still a subject of controversy.
The spin-spin correlation function $\chi_{Ph}({\bf
q},q_{z},\omega)
\equiv \chi^{\prime} + i\chi^{\prime\prime}$, ${\bf q} \equiv
(q_{x},q_{y})$,
usually probed by
NMR, magnetic Raman scattering or, more directly, by
inelastic neutron scattering (INS) experiments,
is still not well understood.
INS measurements on YBa$_2$Cu$_3$O$_{6+x}$ (YBCO) have
been performed by a
number of groups\cite{RossatTranquada,Tranquada92,BourgesRegnault}
and have lead to a consistent picture of the measured physical
susceptibility
$\chi_{Ph}({\bf q},q_{z},\omega)$ for ${\bf
q}$ near the antiferromagnetic (AF) wave vector ${\bf Q} \equiv
(\pi, \pi)$.
It has been found that the dynamical AF spin
fluctuations persist over the whole range of doping ($ 0.4 \leq x
\leq
1$) and the following remarkable features have been reported:

(a) In the superconducting (SC) state
$\chi_{Ph}^{\prime\prime}({\bf Q},q_z,\omega)$ is
limited by an energy gap, $E_g$ (Ref.
\onlinecite{BourgesRegnault}). In
heavily doped regime ($ 0.65 \leq x \leq 0.92$) $E_{g}$
is roughly proportional to
the SC transition temperature $T_c$, that is $E_g\approx
3.4~k_{B}T_{c}$.
In the weakly-doped regime $E_g$ is very small and it is a much
stronger
function of $T_c$ (Ref. \onlinecite{Regnault94}).

(b) In addition, in highly doped YBCO
the spin excitation spectrum is characterized by a
sharp peak at some energy $E_r > E_g$, e.g., for optimally doped
YBa$_2$Cu$_3$O$_{6.92}$, $E_r = 41$~meV (Refs.
\onlinecite{Rossat-Mignod91,Mook93,Fong95}).
For underdoped samples a broad resonance has been observed at
$20 < E_r <40$~meV (Refs. \onlinecite{Tranquada92,Regnault94}).
The resonance excitation is localized in both energy and wave
vector and
has $\chi_{Ph}^{\prime\prime}({\bf Q},q_z,E_r)\propto \sin^2(q_z d
/2)$
modulation,
where $d$ is the distance between two adjacent
CuO$_2$-planes in the unit cell of YBCO.

(c) At $\omega\approx 50$~meV there is a high energy cut-off,
i.e.,
the intensity sharply decreases, regardless of the doping level.

(d) In the normal state
$\chi_{Ph}^{\prime\prime}({\bf Q},q_z,\omega)$
is characterized by a broad feature, spread practically over the
whole energy range, with a characteristic energy
$\omega\sim 30$~meV, approximately independent of the oxygen
concentration.

There has been a number of theoretical proposals for the
explanation of features (a)-(d).
It has been shown by Lu\cite{Lu} that for $d$-wave SC
state $\chi^{\prime\prime}({\bf q},q_{z},\omega)$ should exhibit
anomalous resonance
peaks in ${\bf q}$ space.
Feature (b) has been treated in terms of  spin-flip excitations
across the
SC gap \cite{BSS,CTH,BS,MS,Maki,SPL,FKT,LS,ORM,Mazin95} or
as collective mode excitations in the particle-particle
channel.\cite{DZ}

In this paper we show that the YBCO bilayer structure and
2D band topology, renormalized for
AF interactions, with $d$-wave SC order parameter, leads to novel
magnetic quasi-particle excitations which explain all of the
features
(a)--(d), observed in INS.

We start by calculating
$\chi_{Ph}^{\prime\prime}({\bf q},q_z,\omega)$
for YBCO:
we consider a system of two 2D planes, corresponding to the
Cu$_2$O$_4$ bilayer
in a unit cell of YBCO,
with an effective band dispersion:
$\xi^{(i)}_{{\bf k}}=-2t[\cos (k_x)+\cos (k_y)]-4t^{\prime}\cos
(k_x)\cos (k_y)-2t^{\prime\prime}[\cos (2k_x)+\cos (2k_y)] \pm
t_{\perp} - \mu$. Here index ($i$) corresponds to
the electron bands
with {\em odd} ($o$) and {\em even} ($e$) wave functions with
respect to the mirror plane between
the adjacent planes.
We choose the parameters $t$, $t'$, $t''$, $t_\perp$ and $\mu$
by requiring that the shape and depth of the band and the size of
the Fermi surface (FS)
are approximately equal to those observed in ARPES measurements.
According to Ref. \onlinecite{VealGu}, the {\em even} band is
nearly half filled regardless of the
oxygen content, while at optimal doping the {\em odd} band is
shifted upwards
by $2 t_\perp \approx 100$~meV and thus it is far away from half
filling.
This yields band parameters $t = 115$~meV,
$t^{\prime}/t = -0.2$,
$t^{\prime\prime}/t = 0.25$,
and $\mu=-228$~meV (density of electrons $n^{(o)}=0.55$ and
$n^{(e)}=0.9$).
Although one usually interprets the parameters
$t$, $t'$ and $t''$ in terms of tight-binding overlap
integrals, here they should be regarded as results of our fits to
the
experimental data and as such, they should include some of
the effects of mass renormalization.
The {\em odd} band has a {\em bifurcated saddle point} at ${\bf
q}=(\pi,0)$,
with $\xi^{(o)}_{\pi, 0} = 30$~meV and $\xi^{(o)}_{0.7 \pi, 0} =
10$~meV
(Ref. \onlinecite{OKAnderson}); however, none of our conclusions
are altered if
one assumes an extended Van Hove singularity (VHS) instead.
The FS of the {\em odd} and {\em even} bands are shown in Fig.
\ref{fig:FS}.

In BCS theory the 2D Fermi liquid spin
susceptibility is given by:\cite{Schrieffer,ULee}
\begin{eqnarray}
&&\chi _0^{(ij)}({\bf q},\omega )=
\sum_{{\bf k}}
l^{(ij)2}_{{\bf k}+{\bf q},{\bf k}}E_{-}^{(ij)}({\bf k},{\bf q})
\frac {f(E_{{\bf k}+{\bf q}}^{(i)}) - f(E_{\bf k}^{(j)})}
{E_{-}^{(ij)2} - (\omega - i\Gamma)^2 } \nonumber \\
&& + \sum_{{\bf k}}
p^{(ij)2}_{{\bf k}+{\bf q},{\bf k}}
E_{+}^{(ij)}({\bf k},{\bf q})
\frac {1 - f(E_{{\bf k}+{\bf q}}^{(i)}) - f(E_{\bf k}^{(j)})}
{E_{+}^{(ij)2} - (\omega - i\Gamma)^2 },
\label{chi0}
\end{eqnarray}
where
$E_{\pm}^{(ij)}= E_{{\bf k}}^{(j)}  \pm  E_{{\bf k}+{\bf
q}}^{(i)}$,
$E_{\bf k}^{(i)}=\sqrt{\xi_{{\bf k}}^{(i) 2} + \Delta_{{\bf
k}}^{(i) 2}}$
is the
quasiparticle dispersion law in the SC state,
$f(E_{\bf k}^{(i)}) = 1/[\exp(E_{\bf k}^{(i)}/T) + 1]$ is the
Fermi function and the
coherence factors are given by
\begin{eqnarray}
&&p^{(ij)2}_{{\bf k}+{\bf q},{\bf k}} =
\frac {1}{2}\left(1 - \frac{\xi _{{\bf k}+{\bf q}}^{(i)} \xi
_{{\bf k}}^{(j)}+
\Delta _{{\bf k}+{\bf q}}^{(i)}\Delta _{{\bf k}}^{(j)}}
{E_{{\bf k}+{\bf q}}^{(i)} E_{\bf k}^{(j)}}\right), \nonumber \\
&&l^{(ij)2}_{{\bf k}+{\bf q},{\bf k}} =
\frac {1}{2}\left(1 + \frac{\xi _{{\bf k}+{\bf q}}^{(i)} \xi
_{{\bf k}}^{(j)}+
\Delta _{{\bf k}+{\bf q}}^{(i)}\Delta _{{\bf k}}^{(j)}}
{E_{{\bf k}+{\bf q}}^{(i)} E_{\bf k}^{(j)}}\right).
\label{coherence}
\end{eqnarray}
We assume that the SC gap parameter $\Delta^{(i)}_{\bf k}$ has a
simple form
$\Delta _{\bf k}^{(i)} = \Delta _{max}^{(i)}[\cos (k_x)-\cos
(k_y)]/2$,
although self-consistent numerical solutions of the gap equation
suggest
that $\Delta^{(i)}_{\bf k}$ may have a somewhat different
shape.\cite{BranchCarbotte}
In Eq.\ (\ref{chi0}) $\Gamma$
is the phenomenological damping parameter.
Accounting for the interaction
between the susceptibility bubbles, we calculate
the spin susceptibility renormalized in the
RPA manner:\cite{Maki,SPL,FKT,ULee,Si}
\begin{equation}
\chi^{(i,j)}({\bf q},\omega) = \chi^{(i,j)}_0({\bf q},\omega)/
[1+J^{(i,j)}({\bf q})\chi^{(i,j)}_0({\bf q},\omega)],
\label{chi}
\end{equation}
where  $J^{(e,e)}({\bf q}) = J_{\Vert} + J_{\perp}$,
$J^{(o,o)}({\bf q}) = J_{\Vert} - J_{\perp}$,
$J^{(e,o)}({\bf q}) = J^{(o,e)}({\bf q}) = 0$, $J_\Vert({\bf q})$
is the
in-plane magnetic interaction and $J_{\perp}({\bf q})$
is the interplane magnetic coupling.
We do not take into account the vertex and self-energy corrections
which would further renormalize $\chi^{(i,j)}({\bf q},\omega)$.

For the bilayer case the physical susceptibility, proportional
to the INS measured intensity, is given by
$\sum_{m,n}\exp\{iq_z(z_m - z_n)\} \chi^{(m,n)}({\bf q},\omega)$
(Ref. \onlinecite{Sato88}), where m and n are layer indexes.
If coherence is preserved within, but not
between the bilayers, this expression can be rewritten:
\begin{eqnarray}
&&\chi_{Ph}({\bf q},q_z,\omega) = \nonumber \\
&&2[\chi^{(e,e)}({\bf q},\omega) \cos^2(q_z d/2) +
\chi^{(o,o)}({\bf q},\omega) \sin^2(q_z d/2)].
\label{chiphys}
\end{eqnarray}
It is clear from Eq.\ (\ref{chiphys}) that
the experimentally observed $\sin^2(q_z d/2)$ modulation of the
INS (b) must
originate from the {\em odd} band, provided the coupling between
bilayers is
incoherent.

For fixed ${\bf q}$ the quantity $E_{+}^{(i,i)}({\bf k},{\bf q})$
is a function of 2D vector ${\bf k}$ and we denote its minima and
saddle points by $E_{min}^{(i,i)}({\bf q})$ and
$E_{sp}^{(i,i)}({\bf q})$
respectively. In a 2D system and in the $\Gamma,T \rightarrow 0$
limit
these extrema produce logarithmic singularities in either
$\chi_0^{\prime(i,i)}({\bf q},\omega)$ or
$\chi_0^{\prime\prime(i,i)}({\bf q},\omega)$ (see Eq.\
(\ref{chi0})).
For example,
for $p^{(i,i)2}_{{\bf k}+{\bf q}} \neq 0$,
$\chi_0^{\prime(i,i)}({\bf q},\omega)$
diverges at $\omega \approx E_{min}^{(i,i)}({\bf q})$, while
$\chi_0^{\prime\prime(i,i)}({\bf q},\omega)$ behaves like a
step function.
Alternatively, $\chi_0^{\prime\prime}({\bf q},\omega)$ diverges
at the saddle point energies, $E_{sp}^{(i,i)}({\bf q})$, while
$\chi _0^{\prime (i,i)}({\bf q},\omega)$ has a
``kink".\cite{Mazin95}
Provided that $J_\Vert$ and $J_\perp$ are not too large,
the renormalized susceptibility (\ref{chi})
has poles (resonances) at $\omega$ close to both
$E_{min}^{(i,i)}({\bf q})$
and $E_{sp}^{(i,i)}({\bf q})$.
Physically, these poles in
$\chi^{(i,i)}({\bf q},\omega)$ describe spin-flip quasi-particle
excitations (magnons).
The self-energy corrections shift the position of the poles to the
complex
plane, producing the finite lifetime of the excitations, and more
importantly,
the logarithmic divergences become finite peaks, introducing
a minimal value of $J^{(i,i)}({\bf q})$ for the creation of a
quasi-particle.
For quasi-2D system the resonances become broader due to the weak
interaction between the
bilayers.

Low energy
excitations at ${\bf q}={\bf Q}$ correspond to particles and holes
near
the intercepts of the FS
and the magnetic Brillouin zone (e.g., points A and B in Fig.\
\ref{fig:FS}).
The SC gap at these points $\Delta^{(o)}_{\bf k_A}$
($=\Delta^{(o)}_{\bf k_B}$) is close to
$\Delta_{max}^{(o)}$, and thus
the {\em lowest} energy excitation [feature (a) in INS]
is $E_g = \Delta^{(o)}_{\bf k_A} + \Delta^{(o)}_{\bf k_B} =
E_{min}^{(o,o)}({\bf Q})$.
On the other hand, holes near
VHSs and particles above the SC gap produce $E_{sp}^{(o,o)}({\bf
q})$
resonances. At ${\bf q}={\bf Q}$ this yields the
excitation energy $E_r=E_{sp}^{(o,o)}({\bf Q})=\Delta^{(o)}_{\bf
k_D} + E^{(o)}_{\bf
k_C}$, where C and D are points depicted in Fig.\ \ref{fig:FS}
[feature (b)].
In Fig.\ \ref{fig:dispersion} we present the energy dispersion of
the
quasi-particles, originating from both $E_{min}^{(o,o)}({\bf q})$
and
$E_{sp}^{(o,o)}({\bf q})$,
for ${\bf q}$ along the high symmetry lines leading to ${\bf Q}$,
and assuming $\Delta^{(o)}_{max}=15$~meV.
Away from ${\bf Q}$ the excitations clearly
split into several bands, whose dispersion depends on the shape of
VHSs,
and $E_{sp}^{(o,o)}({\bf q})$ reaches approximately 0.3~eV near
the
magnetic Brillouin zone boundary.
Of course, the excitations (resonances)
shown in Fig.\ \ref{fig:dispersion} may have
very different lifetimes as well as intensities, depending not
only
on the FS geometry, but also on the appropriate coherence factors.

Returning to the INS results, both $E_g$ and $E_r$
can be observed experimentally {only} if damping is sufficiently
small.
We calculated the single electron (hole) magnetic relaxation rate
$\tau_{\bf
k}^{-1}$ (see Eq.\ (9b) in Ref. \onlinecite{Si}) and
found that in the SC state the relaxation is negligible for {\bf
k} near
$(\pi,0)$ and energies below $\sim 2.4 \Delta_{max}$,
in contrast with ordinary superconductors where $\tau_{\bf
k}^{-1}(\omega,T)$
vanishes for $\omega < 3\Delta$ (Ref. \onlinecite{Schrieffer}).
As a result, any spin-flip excitation in
$\chi^{\prime\prime(i,i)}({\bf Q},\omega)$,
will not be damped for $\omega$ smaller than approximately
$3.4\Delta_{max}$
even in $d$-wave superconductors.
Thus, both $E_g$ and $E_r=  \Delta^{(o)}_{\bf k_D} + E^{(o)}_{\bf
k_C}$
{\em should} be observed experimentally.
On the other hand, in the normal state the $E_r$ excitation
becomes
heavily damped and is sometimes referred to as the {\em
spin-pseudogap}.
We note that for the {\em even} band the saddle point energy lies
far below the FS and the corresponding
resonance excitation is overdamped even in the SC state and
contributes only to a structureless background.

The temperature dependence of the calculated {\em odd} band
susceptibility $\chi^{\prime\prime(o,o)}({\bf Q},\omega)$ is
presented in
Fig.~\ref{fig:T-dop-dependence}a.
We assume a BCS temperature dependence of
the gap parameter, although, admittedly, $\Delta^{(o)}_{max}(T)$
can be regarded
as a fitting parameter in order to match the experimental INS
spectra.\cite{Regnault94,Rossat-Mignod91,Mook93,Fong95}
Single band fits to NMR Knight shift suggest a larger value of SC
gap parameter\cite{Pines} than that we use here
($\Delta^{(o)}_{max} =
15$~meV at $T = 0$). In the bilayer case only {\em
even} band contributes to $\chi^\prime_{{Ph}}(0,0,0)$
[Ref.\onlinecite{MM}], i.e., $\Delta^{(e)}_{max}$ may be larger
than $\Delta^{(o)}_{max}$ (Ref. \onlinecite{ULee}). In addition,
a different form of $\Delta^{(i)}_{\bf k}$ may alter
$\Delta^{(i)}_{max}$ necessary to fit experiments.
We assume $J^{(o,o)}({\bf Q}) = -120$~meV.
The calculated spectra reproduce all of the observed INS features
in the SC state: the gap
originating from the SC gap parameter (a); the sharp
resonance peak at 41~meV at low temperatures,
due to spin-flip quasi-particle excitations
related to the VHS in the band structure (b);
the sharp drop of the signal above 45~meV, also due to the band
topology and
the effect of renormalization (c).
The overall enhancement of
$\chi^{\prime\prime(o,o)}({\bf Q},\omega)$
is due to
the non-vanishing coherence factor $p^{(o,o)2}_{{\bf k_A},{\bf
k_B}}$,
present in $d$-wave superconductors\cite{Lu,MS}
($\Delta^{(i)}_{\bf k_A}\Delta ^{(i)}_{\bf k_B} < 0 $ ).
At temperatures above $T_c$ the magnetic scattering is depressed.
However, in
agreement with the experimental data (d), a broad peak at
$\omega\approx
20-30$~meV remains ({\em spin-pseudogap}).
Its origin is the same as that of
$E_r$ resonance in the SC state (VHS);
its position is shifted due to the SC gap vanishing.

The doping dependence of
$\chi^{\prime\prime(o,o)}({\bf Q},\omega)$ is shown in
Fig.~\ref{fig:T-dop-dependence}b. For the underdoped case
we applied the same criteria as for the optimally
doped, i.e., the band parameters are chosen in accordance with
ARPES measurements \cite{VealGu} which show that the VHS at
$(\pi,0)$
remains approximately at the same energy below the FS, yet the
filling
is, obviously, much smaller. Therefore, in order to
fit the experimental data
we assumed, for the underdoped case, that $t=106$~meV,
$t'/t=-0.18$, $t''/t=
0.18$ and $\mu=-125$~meV ($n^{(o)}=0.7$), with
$\Delta^{(o)}_{max}=5$~meV and
$J^{(o,o)}({\bf Q})=-200$~meV,
and for the overdoped case we assumed
$t=124$~meV, $t'/t=-0.22$, $t''/t=0.24$ and $\mu = -204$~meV
($n^{(o)}=0.5$) with
$\Delta^{(o)}_{max}$ equal to that in the optimally doped case and
$J^{(o,o)}({\bf Q})=0$.
The calculated result is qualitatively similar
to the experimental one.\cite{Regnault94}
Most importantly, since the value of $\Delta^{(o)}_{max}$ is
smaller for
the underdoped materials, but the distance from the VHS
at ${\bf q} = (\pi,0)$ to the FS remains the same as in the
optimally doped case,\cite{VealGu} the resonance $E_r$ must become
broad
at some doping level when this distance is
approximately equal to $2.4\Delta^{(o)}_{max}$,
and this is what the experiment suggests.

In conclusion, we have shown that the Van Hove singularities,
observed in ARPES at approximately 30~meV below the Fermi level,
lead to a novel, well-defined magnetic quasi-particle excitations
in the SC state, for quasi-2D systems and $d$-wave gap symmetry.
Our model calculations of $\chi_{Ph}^{\prime\prime}({\bf
Q},q_z,\omega)$ are
consistent with {\em all} major features observed in INS
experiments.
The model suggests a relatively small
value for $\Delta^{(o)}_{max} \approx 1.9 k_BT_c$, and a
relatively weak
Heisenberg interaction for the {\em odd} band of optimally doped
YBCO.
In addition, our results show that
the comparison of INS data and other magnetic probes, such as NMR,
where
both {\em odd} and {\em even}
bands must be taken into account, may not be as straight-forward
as previously believed.

We are indebted to V.\ Barzykin, A.\ J.\ Leggett, D.\ Pines and
O.\ T.\ Valls
for valuable discussions.
This work has been supported in part
by NSF cooperative agreement DMR 91-20000 through the STCS.

\begin{figure}
\caption{Fermi surfaces of odd (solid line) and even bands (dashed
line)
in YBCO. The parameters
are given in the text.
The dash-dotted line shows the magnetic Brillouin zone. Points
A and B (C and D) can be connected by the AF wave vector ${\bf
Q}\equiv
(\pi,\pi)$ and
correspond to $E_g$ ($E_r$) resonances observed in INS
measurements
(see text).}
\label{fig:FS}
\end{figure}

\begin{figure}
\caption{
Magnetic quasi-particle dispersion for {\bf q} along lines of high
symmetry,
in the vicinity of {\bf Q}. The values are obtained from
the minima (triangles) and saddle points
(solid circles) of the particle-hole energy $E_+({\bf k,q})$ [see
Eq.\ (1)]
at fixed momentum transfer ${\bf q}$.
}
\label{fig:dispersion}
\end{figure}

\begin{figure}
\caption{The {\em odd} band susceptibility
$\chi^{\prime\prime(o,o)}({\bf Q},\omega)$, defined in Eq.\ (3)
and proportional
to the INS measured intensity, as a function of
energy.
No finite lifetime effects
are included (in both normal and SC state $\Gamma=0.5$~meV).
$\chi^{\prime\prime(o,o)}({\bf Q},\omega)$ is gaped in the SC
state and has a peak
at $E_{sp}^{(o,o)}({\bf q})$ (see text) due to the VHS below the
FS.
(a) The results for optimally doped YBCO corresponding to
$T=0$, 80, 100, 200, 300~K (top to bottom).
(b) The results for different doping levels and at $T=0$~K.}
\label{fig:T-dop-dependence}
\end{figure}

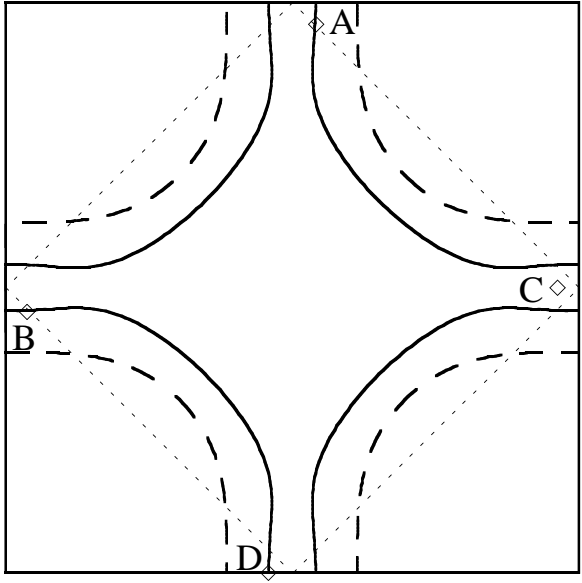
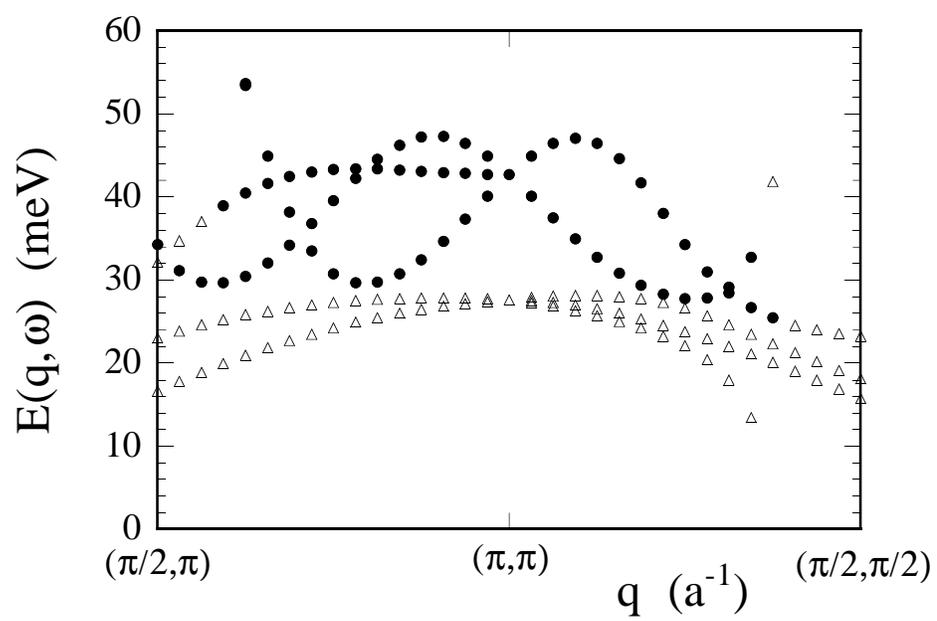
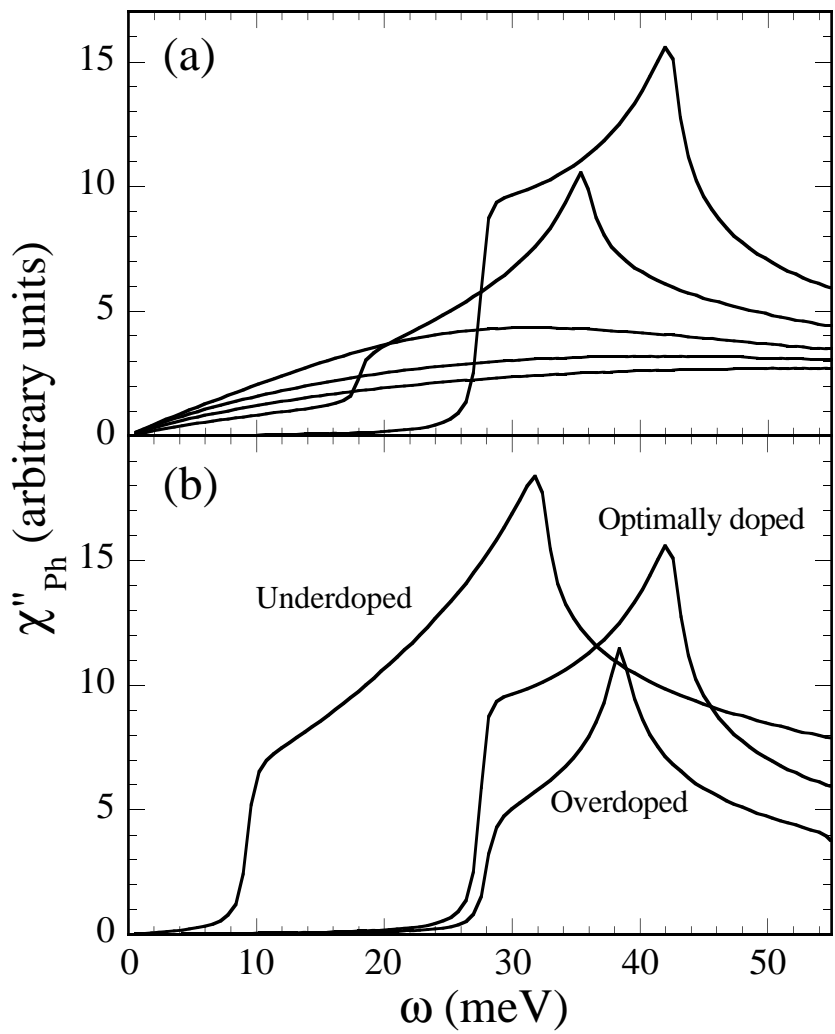
\end{document}